\documentclass[aps,pre,twocolumn,superscriptaddress]{revtex4-1}
\usepackage{amsmath}
\usepackage{amssymb}
\usepackage{graphicx}
\usepackage{subfigure}

\begin{document}

%\preprint{}

\title{Mean-field model for the Curie-Weiss temperature dependence of coherence length in metallic liquids}


\author{Charles K. C. Lieou}
\email[]{clieou@utk.edu}
%\thanks{}
\affiliation{Department of Materials Science and Engineering, University of Tennessee, Knoxville, TN 37996, USA}
%\affiliation{Earth and Environmental Sciences, Los Alamos National Laboratory, Los Alamos, NM 87544, USA}
\author{Takeshi Egami}
\affiliation{Department of Materials Science and Engineering, University of Tennessee, Knoxville, TN 37996, USA}
\affiliation{Department of Physics and Astronomy, University of Tennessee, Knoxville, TN 37996, USA}
\affiliation{Materials Science and Technology Division, Oak Ridge National Laboratory, Oak Ridge, TN 37831, USA}

\date{\today}

\begin{abstract}
The coherence length of the medium-range order (MRO) in metallic liquids is known to display a Curie-Weiss temperature dependence; its inverse is linearly related to temperature, and when extrapolated from temperatures above the glass transition, the coherence length diverges at a negative temperature with a critical exponent of unity. We propose a mean-field pseudospin model that explains this behavior. Specifically, we model the atoms and their local environment as Ising spins with antiferromagnetic exchange interactions. We further superimpose an exchange interaction between dynamical heterogeneities, or clusters of atoms undergoing cooperative motion. The coherence length in the metallic liquid is thus the correlation length between dynamical heterogeneities. Our results reaffirm the idea that the MRO coherence length is a measure of point-to-set correlations, and that local frustrations in the interatomic interactions are prominent in metallic liquids.
\end{abstract}

%\keywords{}

\maketitle


\section{Introduction}\label{sec:1}

The structure of a metallic liquid and glass, assumed to be isotropic in what follows, is measured by the atomic pair distribution function (PDF) \cite{hansen_2013,egami_2020},
\begin{equation}\label{eq:gr}
 g(r) = \dfrac{1}{4 \pi N \rho_0 r} \sum_{i,j} \left\langle \delta( r - | \mathbf{r}_i - \mathbf{r}_j | ) \right\rangle ,
\end{equation}
and its Fourier transform, termed the structure function
\begin{equation}\label{eq:Sq}
 S(q) = 1 + \dfrac{4 \pi \rho_0}{q} \int_0^{\infty} dr \left[ g(r) - 1 \right] r \sin (q r) .
\end{equation}
In Eq.~\eqref{eq:gr}, $N$ is the total number of atoms, $\rho_0$ is the number density of atoms, $\mathbf{r}_i$ is the position of the atom labeled $i$, and $\left\langle \cdots \right\rangle$ denotes the thermal average. The medium-range order (MRO) of a metallic liquid is measured by the height of the first peak of the structure factor, $S(Q_{\text{MRO}})$, and the coherence length $\xi$, which defines the decay of the envelope of the pair distribution function beyond the first peak:
\begin{equation}
 G(r) \equiv 4 \pi r \rho_0 \left[ g(r) - 1 \right] = G_0(r) \exp \left( - \dfrac{r}{\xi} \right) ,
\end{equation}
where $G_0(r)$ specifies the ideal glass state with infinite-range density correlations. Recent experiments and simulations~\cite{ryu_2019} confirm that the quantity $S(Q_{\text{MRO}}) - 1$, and therefore the coherence length $\xi$, follow Curie-Weiss temperature dependence,
\begin{equation}
 \xi \propto a (\theta + \theta_c)^{-1} ; \quad \text{for $\theta > \theta_N$},
\end{equation}
in the high-temperature regime, above some temperature $\theta_N$. When extrapolated from temperatures $\theta \equiv k_B T$ above the glass transition, both quantities are seen to diverge at a \textit{negative} Curie temperature $- \theta_{IG} < 0$. The glass transition temperature $\theta_g$ is indicated by a cusp in the quantities $\xi$ and $S(Q_{\text{MRO}}) - 1$. Concomitant with the rapid increase of both quantities is the increase in viscosity of the metallic liquid. Indeed, it is suggested in \cite{ryu_2019} that the viscosity $\eta_a$ increases with decreasing temperature as
\begin{equation}\label{eq:eta_a}
 \eta_a (\theta) = \eta_0 \exp \left( \dfrac{E_a (\theta)}{\theta} \right) ,
\end{equation}
where $E_a (\theta)$, the activation energy, scales with the temperature-dependent coherence length $\xi(\theta)$ as $E_a (\theta) \propto \left( \xi(\theta) \right)^3$, the typical size of a cooperatively rearranging cluster of atoms \cite{ryu_2019}. These results are independent of the specific form of the interatomic potential, suggesting that some fundamental, medium-range mechanisms are responsible for the cooperative behavior of atoms in supercooled liquids. Elucidation of such mechanisms may shed important light on the nature of the coherence length, as well as the physics of the glass transition itself.

Recent simulations by Tanaka and coworkers \cite{tanaka_2010,kawasaki_2011,tanaka_2012} and theoretical analyses of these results \cite{langer_2013,langer_2014} point to the physical picture of glass-forming liquids containing a population of twofold degenerate clusters of atoms with local topological order, that can be described by pseudospin models that are often invoked to explain glassy behavior. Indeed, the idea of using two-state spin variables to represent structural features of glass-forming liquids is not new; there may be some connections between structural glasses and spin glasses~\cite{binder_1986}, However, the relevance of popular spin-glass models such as the Sherrington-Kirkpatrick (SK) model \cite{sherrington_1975,kirkpatrick_1978,parisi_1979,parisi_1980,parisi_1983} and the random-first-order transition theory (RFOT) \cite{xia_2000,xia_2001,lubchenko_2007}, which embody infinite-range interactions between all spins, to structural glasses with short-range atomic and molecular interactions, may be called into question. In addition, a straightforward mean-field analysis of typical spin models necessarily gives a correlation-length critical exponent of $\nu = 1/2$ \cite{binder_1986}, contrary to $\nu = 1$ seen in the medium-range ordering behavior of glass-forming liquids \cite{ryu_2019}.

The crux of the matter, in our view, is that the spin-spin correlation length refers to a short-range two-body correlation that is decoupled from the MRO probed by studies such as Ref.~\cite{ryu_2019}. Indeed, MRO represents the correlation between an atom and coarse-grained density fluctuations at some distance $r$ away, or point-to-set correlations \cite{berthier_2012}. Such density fluctuations are represented by local clusters of atoms; the cluster size is determined by short-range two-point correlations. In other words, the coherence length $\xi$ represents the range over which these clusters are correlated, and is certainly different from the two-point or spin-spin correlation length.

The present paper makes these ideas quantitatively explicit by building upon the spin-like binary cluster model proposed in \cite{langer_2013,langer_2014}. In Sec.~\ref{sec:2} we present a pseudospin model that takes into account both individual atoms and the topologically correlated atoms, termed dynamical heterogeneities, where rearrangement may occur. Then, in Sec.~\ref{sec:3}, we use statistical thermodynamics to derive the critical properties of this model, and show that a coherence length exponent of $\nu = 1$ naturally emerges from generic considerations. We make connections to simulations and experiments in Sec.~\ref{sec:4} and conclude with a summary and outstanding questions in Sec.~\ref{sec:5}.

\section{Ising model of pseudospins and dynamical heterogeneities: internal state variables}
\label{sec:2}

Our physical picture of a metallic liquid consists of short-range interactions between atoms, as well as clusters of atoms where cooperative motion may occur. The latter are termed dynamical heterogeneities, shear transformation zones, and binary clusters elsewhere in the literature \cite{falk_1998,falk_2011,langer_2013,langer_2014,lieou_2014}.

Two features should emerge from this model: (1) a coherence length exponent of $\nu = 1$; and (2) a glass transition temperature $\theta_g$ at which the first peak of the structure factor $S(Q_{\text{MRO}})$, which measures medium-range order, and the coherence length $\xi \propto S(Q_{\text{MRO}}) - 1$, display a cusp as a function of temperature. This cusp is analogous to that seen in an antiferromagnet which calls for at least two order parameters. To this end, we introduce two order parameters $m$ and $\eta$, which are analogous to the total and staggered magnetization per unit volume. Physically, one can think of this as introducing two measures of the local atomic environment (e.g., atomic-level stresses \cite{egami_1980}) $A$ and $B$, and in the spirit of spin models, assign a value of $+1$ or $-1$ to each atom for each of these measures. Thus, if $N$ is the total, extensive number of atoms in the system, and $N_+^A$, $N_-^A$, $N_+^B$, $N_-^B$ are the number of atoms in states $+1$ and $-1$ for each of the two measures $A$ and $B$, then $N = N_+^A + N_-^A = N_+^B + N_-^B$, and we can write
\begin{equation}
 m = \dfrac{N_+^A - N_-^A}{N}; \quad \eta = \dfrac{N_+^B - N_-^B}{N} ,
\end{equation}
which give rise to interaction energies proportional to $- J m^2$ and $- K \eta^2$, respectively, in the mean-field description, with $J < 0$ and $K > 0$. We also include an interaction term proportional to $m^2 \eta^2$ which, as we shall see, is needed to account for the cusp in the correlation length at the glass transition temperature. Note that we do not attempt to invoke the replica method here which, while rather typical for a mathematical description of metastable states in glasses, give results with similar mathematical structure in the mean-field description.

In addition to $m$ and $\eta$, we introduce the density of dynamical heterogeneities $\Lambda$ and orientational bias $M$, defined in terms of the number of dynamical heterogeneities in each of the two states $N_+$ and $N_-$, that can be thought of as their orientations with respect to some direction in space, and the total number of sites $N_S$ where a dynamical heterogeneity can appear:
\begin{equation}
 \Lambda = \dfrac{N_+ + N_-}{N_S}; \quad M = \dfrac{N_+ - N_-}{N_+ + N_-} .
\end{equation}
In the spin language, one can think of dynamical heterogeneities as block spins. Note that $N_S$ decreases with decreasing temperature which increases the cooperativity between atoms and hence the typical size $\xi_S$ of a dynamical heterogeneity; we expect $N_S \propto \xi_S^{-3}$. Dynamical heterogeneities are the only contributors to plastic strain, but we do not consider plastic strain in the metallic liquid that is not undergoing deformation.

\section{Statistical thermodynamics: energy, entropy, and critical behavior}
\label{sec:3}

With these ingredients, we can write the total energy of the system as (see also, for example, \cite{langer_2013,lieou_2014}):
\begin{eqnarray}\label{eq:U}
 \nonumber U(\Lambda, M, m, \eta) &=& N_S \Lambda e_Z - \dfrac{N_S}{2} J' \Lambda M^2 \\ & & - \dfrac{N}{2} \left[ J m^2 (1 + \alpha \eta^2) + K \eta^2 \right] .
\end{eqnarray}
Here, $J' < 0$ is a negative interaction constant which reflects the intuition that in a zero-stress environment, it is energetically unfavorable for dynamical heterogeneities to display directional bias. The quantity $\alpha > 0$ is a small constant. The formation energy of a dynamical heterogeneity, denoted by $e_Z$, is expected to scale as its size $\xi_S^3$. Moreover, the entropy can be computed by simple counting. For example, the contributions from the local atomic environment equals the logarithm of the total number of microstates
\begin{equation}
 W = \dfrac{N!}{N_+^A ! N_-^A !} \dfrac{N!}{N_+^B ! N_-^B !} .
\end{equation}
Upon taking the Stirling approximation, and doing the same for the contributions from dynamical heterogeneities, the result is
\begin{eqnarray}\label{eq:S}
 \nonumber S(\Lambda, M, m, \eta) &=& N_S S_0 (\Lambda) + N_S \Lambda \psi(M) \\ & & + N \left[ \psi(m) + \psi(\eta) \right ], 
\end{eqnarray}
where
\begin{eqnarray}
 S_0 (\Lambda) &=& - \Lambda \ln \Lambda + \Lambda ; \\
 \psi(m) &=& \ln 2 - \dfrac{1+m}{2} \ln (1 + m) - \dfrac{1-m}{2} \ln (1 - m) . ~~~~~
\end{eqnarray}

One proceeds by looking for minima in the free energy landscape, with the free energy being
\begin{equation}
 F = U - \chi S ,
\end{equation}
by computing the derivatives
\begin{eqnarray}
 \dfrac{\partial F}{\partial \Lambda} &=& N_S \left[ e_Z - \dfrac{J'}{2} M^2 + \chi \left( \ln \Lambda - \psi(M) \right) \right] ;\\
 \dfrac{\partial F}{\partial M} &=& - N_S \Lambda \left(J M - \chi \tanh^{-1} M \right) ; \\
 \dfrac{\partial F}{\partial m} &=& - N \left[ J m (1 + \alpha \eta^2) - \chi \tanh^{-1} m \right] ; \\
 \dfrac{\partial F}{\partial \eta} &=& - N \left[ (K + J \alpha m^2) \eta - \chi \tanh^{-1} \eta \right] ,
\end{eqnarray}
and setting them equal to zero. Here, we have used the \textit{effective} temperature $\chi$ instead of the true thermal temperature $\theta$ for greater generality \cite[e.g.,][]{falk_2011,langer_2013,langer_2014}; $\chi = \theta$ in thermal equilibrium but $\chi > \theta$ if the material is out of equilibrium. The order parameters corresponding to the free energy minimum are thus given by
\begin{eqnarray}
 \label{eq:Lambda} \Lambda &=& \exp \left[ - \dfrac{e_Z - (J/2) M^2}{\chi} + \psi(M) \right] \approx 2 \exp \left( - \dfrac{e_Z}{\chi} \right) ; ~~~~~\\
 \label{eq:Ms} M &=& \tanh \left( \dfrac{J' M}{\chi} \right) ; \\
 \label{eq:m} m &=& \tanh \left[ \dfrac{J (1 + \alpha \eta^2) m}{\chi} \right] ; \\
 \label{eq:eta} \eta &=& \tanh \left[ \dfrac{(K + J \alpha m^2) \eta}{\chi} \right].
\end{eqnarray}
With $J' < 0$, $J < 0$, it is clear that we require $M = 0$, $m = 0$; this does not mean, however, that there can be no spatial fluctuations in these quantities. We first note that with $m = 0$, it follows immediately from Eq.~\eqref{eq:eta} that there is a ``phase transition'' in the ``staggered magnetization'' $\eta$ at the ``Neel temperature'' $\theta_N \equiv K$; $\eta$ increases continuously from zero below $\theta_N$:
\begin{equation}\label{eq:eta_-}
 \eta^2 = \dfrac{3 (\theta_N / \chi - 1)}{(\theta_N / \chi)^3} , \quad \text{for $\chi < \theta_N$ and $\eta \ll 1$.}
\end{equation}
Above $\chi = \theta_N$, the stable solution is $\eta = 0$. Then, by including fluctuations $\sim \nabla^2 m$ on the right-hand side of Eq.~\eqref{eq:m}, and taking the Fourier transform, we can show that when extrapolated from the equilibrium, high-temperature regime ($\chi = \theta > \theta_N$), the correlation length $\xi_S$ in the local atomic environment, measured by the order parameter $m$, diverges at the negative temperature $- \theta_c \equiv J$ with exponent $\nu = 1/2$:
\begin{equation}
 \xi_S \propto a (\theta + \theta_c)^{-1/2} ; \quad \text{for $\theta > \theta_N$},
\end{equation}
where $a$ is the atomic diameter. The more general result, which holds below $\theta_N$, is
\begin{equation}\label{eq:xi_S}
 \xi_S \propto a [\chi + (1 + \alpha \eta^2) \theta_c]^{-1/2} .
\end{equation}

Meanwhile, the correlations between dynamical heterogeneities are measured by the correlation or coherence length $\xi$, given \textit{in units of the size of a dynamical heterogeneity} or correlation length for the local atomic environment $\xi_S$:
\begin{equation}\label{eq:xi}
 \xi \propto \xi_S (\chi + \theta_c')^{-1/2} ,
\end{equation}
where the critical temperature is now $- \theta_c' = J'$. We now combine Eqs.~\eqref{eq:xi_S} and \eqref{eq:xi}, and assume further that $J = J'$, which appears to be physically reasonable. This is because both repulsive interactions originate from the atomic potential, and that at high-enough temperatures, where the coherence length equals the atomic size, there is no essential difference between interatomic and dynamical heterogeneity interactions. Then,
\begin{equation}\label{eq:xi_f}
 \xi \propto \dfrac{a}{(\chi + \theta_c)^{1/2} [\chi + (1 + \alpha \eta^2) \theta_c]^{1/2}} \equiv a f (\chi).
\end{equation}
In particular, when extrapolated from the equilibrium, high-temperature regime where $\eta = 0$, one recovers the result
\begin{equation}\label{eq:xi_h}
 \xi \propto \dfrac{a}{\theta + \theta_c} ; \quad \text{for $\theta > \theta_N$} ,
\end{equation}
observed in \cite{ryu_2019}. Thus one can identify the ``Curie temperature'' as $\theta_{IG} = \theta_c$.

\section{Results and Discussion}
\label{sec:4}

\begin{figure}
\begin{center}
\subfigure{\includegraphics[scale=0.5]{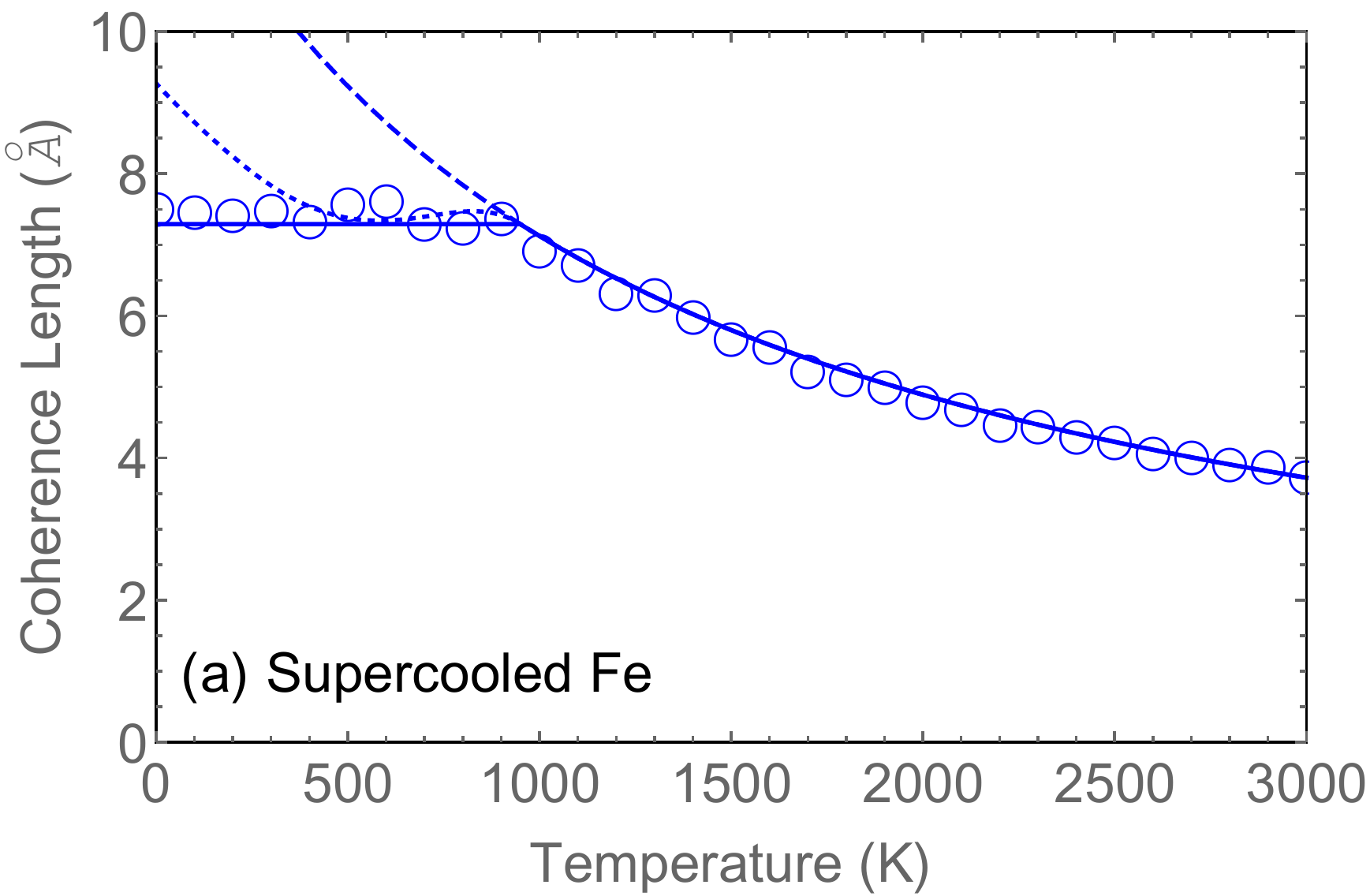}}
\subfigure{\includegraphics[scale=0.5]{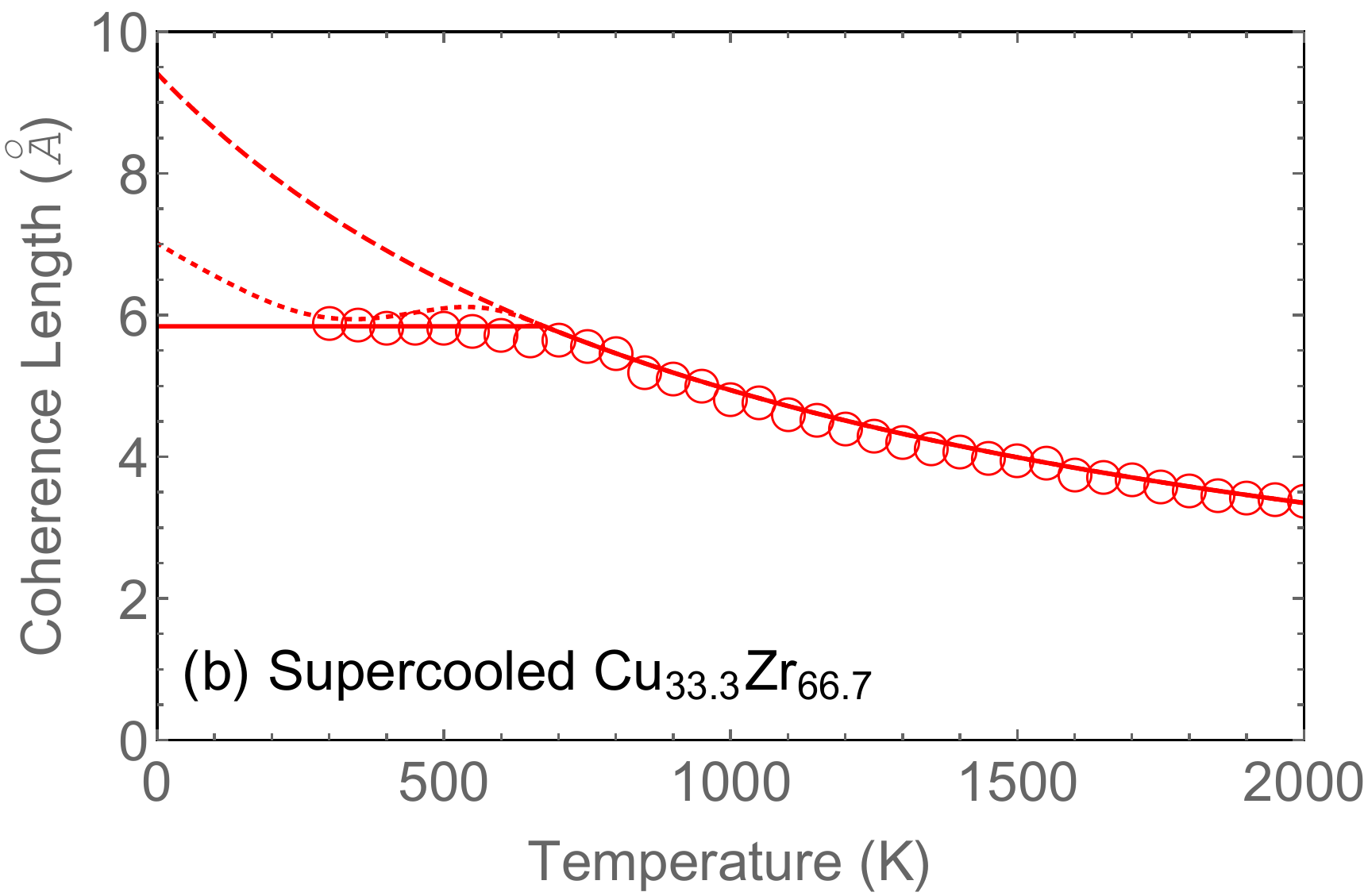}}
\caption{\label{fig:xi}Coherence length of supercooled (a) Fe and (b) Cu$_{33.3}$Zr$_{66.7}$ as functions of temperature. The solid curve represents the coherence length given by Eq.~\eqref{eq:xi_f}, with $\chi = \theta$ for $\theta > \theta_N$ and $\chi = \theta_N$ for $\theta < \theta_N$. The dotted curve represents the thermal equilibrium results given by Eq.~\eqref{eq:xi_f}, with $\chi = \theta$ at all temperatures; the dashed curve represents the high-temperature extrapolation given by Eq.~\eqref{eq:xi_h}. The open circles represent simulation data using LAMMPS.}
\end{center}
\end{figure}

Figure \ref{fig:xi} shows the coherence length of supercooled Fe and Cu$_{33.3}$Zr$_{66.7}$ over a range of temperatures up to 3000 K for Fe, and 2000 K for Cu$_{33.3}$Zr$_{66.7}$. The data points were computed from LAMMPS simulation data; details of the data and simulations are described in \cite{ryu_2021,egami_2021}. The theoretical curve is computed from Eqs.~\eqref{eq:xi_f} and \eqref{eq:xi_h}. In units of Kelvin, $J = - \theta_{IG} = - 1190$ K, $K = \theta_g = 950$ K for Fe, and $J = -1105 $ K, $K = 675$ K for Cu$_{33.3}$Zr$_{66.7}$, with $a$ rescaled accordingly to fit the coherence length data, and $\alpha = 0.8$. In both cases the model metallic liquid has been supercooled and is therefore out of equilibrium with respect to crystallization, but at the cooling rates used in simulation ($10^{11}$ K/s) it is in equilibrium within the liquid phase. However, at low temperatures it undergoes the glass transition and goes out of equilibrium, so that disorder and atomic as well as point-to-set correlations are frozen in, as is suggested by the simulation data. As such, we have used $\chi = \theta$ for $\theta > \theta_N = \theta_g$ above the glass transition, and $\chi = \theta_N = \theta_g$ below the glass transition.

Our derivation leading to Eq.~\eqref{eq:xi_f} for the MRO coherence length -- which involves the notion of ``block spins'' -- is motivated by the idea that the coherence length is really a measure of point-to-set correlations, as opposed to short-range point-to-point correlations between pseudospins. The difference between the SRO and the MRO was clearly demonstrated by recent experimental and simulation results of the variation in the PDF with temperature \cite{ryu_2021}. Whereas the height of the first peak of the PDF, which describes the SRO, changes smoothly with temperature through the glass transition at $\theta_g$, the height of the third peak, which is a part of the MRO, shows significant break in the slope at $\theta_g$. The first peak of the PDF reflects the positions of 12 to 14 near-neighbor atoms. In comparison, the third peak, whose width is about 0.1 nm, represents a few hundred atoms. Therefore it no longer describes the position of each atom, but describes coarse-grained density fluctuations \cite{egami_2020}. A convenient parameter to describe the local density fluctuations is the atomic-level pressure
\cite{egami_1980},
\begin{equation}\label{eq:pressure}
 p_i = \sum_j \dfrac{1}{V_i} \mathbf{f}_{i, j} \cdot \mathbf{r}_{i, j} ,
\end{equation}
where $V_i$ is the volume of the atom $i$, $\mathbf{f}_{i, j}$ is the two-body force and $\mathbf{r}_{i, j}$ is the separation, between an atom $i$ and its neighbors $j$. Because of the summation over the neighbors it is a quantity already averaged over the local neighborhood. While the first moment of the atomic-level pressure is zero, the second moment is proportional to temperature above the glass transition temperature, and shows a change in the slope, unlike the first peak of the PDF \cite{chen_1988}. It was recently argued in \cite{egami_2021} that MRO emerges from the atomic-level pressure fluctuations. This is not inconsistent with our conjecture that MRO emerges from the cooperative behavior in dynamical heterogeneities, for dynamical heterogeneities themselves are a manifestation of local density fluctuations and, being sites at which atomic rearrangements occur, must accommodate local pressure fluctuations. The exact relationship between local density fluctuations as described by Eq.~\eqref{eq:pressure} and the pseudospin variable is beyond the scope of the present paper, but will be an interesting subject for future studies.

%\begin{figure}
%\includegraphics[width=.4\textwidth]{fig_R0.pdf}
%\caption{\label{fig:R0}Variation of the STZ event rate $R_0$ (see text for its definition) with pressure. $1/R_0$ appears to be scale linearly with pressure $p$, suggesting slower slips, rearrangements, and glassy dynamics, as one goes through the jamming transition.}
%\end{figure}

\section{Concluding Remarks}
\label{sec:5}

In this work we have demonstrated that a coherence length critical exponent of $\nu = 1$ is emblematic of point-to-set correlations which itself is a hallmark of medium-range order. Specifically, the Curie-Weiss behavior of the coherence length $\xi$ results from the cooperative behavior of topologically ordered clusters of atoms, or dynamical heterogeneities, whose size $\xi_S$ is determined by the point-to-point correlations in the glass-forming liquid, and grows with decreasing temperature. Our model shows, in addition, that the cusps in the quantities $\xi$ and $S(Q_{\text{MRO}}) - 1$ as functions of temperature are direct consequences of the frustration in the atom-atom interactions, analogous to antiferromagnetic ordering. A major implication of our model results is that cooperative behavior of dynamical heterogeneities is a key contributor to the rapidly increasing viscosity with decreasing temperature as one approaches the glass transition, in accordance with Eq.~\eqref{eq:eta_a}. While the glass transition itself emerges from the frustration in the interactions, the clustering behavior of atoms do play a prominent role here.

We note that in the present work, we did not attempt to model structural disorder with conventional tools typical in the study of spin glasses, such as the replica trick~\cite{binder_1986,parisi_1979,parisi_1980,parisi_1983}. Rather, our point is to demonstrate the essential ingredients for the diverging coherence length and the Curie-Weiss behavior. To that end, a few coarse-grained internal state variables along with a mean-field description would suffice. This has enabled us to extract the coherence length directly from the Ising pseudospins model framework instead of resorting to some hypothetical dynamic mechanism from which one deduces a correlation length, as has been done in RFOT \cite{xia_2000,xia_2001,lubchenko_2007}. On the other hand, it is known that below the glass transition, replica symmetry is broken and the glass-forming material sits in one of many degenerate free energy minima \cite{parisi_1979,parisi_1980,parisi_1983}. The extent to which the simplified analog $\eta$ of the ``staggered magnetization'' in the present Ising model of pseudospins and dynamical heterogeneities describe, or fail to describe, the slow dynamics of heterogeneities below the glass transition, will be an important subject of future investigation.

\begin{acknowledgments}
The authors would like to thank James Langer for illuminating discussions. This work was supported by the US Department of Energy, Office of Science, Basic Energy Sciences, Materials Sciences and Engineering Division.
\end{acknowledgments}

\bibliography{pre_ising_04}

\end{document}